\documentstyle[jltp,11pt,twoside,amssymb]{article}

\title{Acoustic Properties of Amorphous Solids at Very Low Temperatures:
The Quest for Interacting Tunneling States\footnote{Published in the
special issue ``Pushing Physics at Low Temperatures" in honor of Prof. F.
de la Cruz. J. Low Temp. Phys. {\bf 135}, 27 (2004).}}

\author{Pablo Esquinazi, Miguel A.
Ramos$^+$ and Reinhard K\"onig$^*$}

\address{Abteilung Supraleitung und Magnetismus,
Universit\"at Leipzig, Linn\'estrasse 5,\\ D-04103 Leipzig, Germany\\
$^+$Departamento de F\'isica de la Materia Condensada, C-III, Instituto
Nicol\'as\\ Cabrera, Universidad Aut\'onoma de Madrid, E-28049 Madrid,
Spain \\ $^*$Leibniz-Institute for Neurobiology, Special Lab Non-Invasive
Brain Imaging,\\ Brenneckestrasse 6, D-39118 Magdeburg, Germany}

\runninghead{P. Esquinazi, M. Ramos, and R. K\"onig}{Acoustic Properties
of Amorphous Solids}
\input psfig
\begin{document}

\begin{abstract}
We discuss the strain dependence of the acoustic properties of amorphous
metals in both normal and superconducting states, in the temperature range
0.1~mK $ \le T \le 1~$K. A crossover is found when the strain energy is of
the order of the effective interaction energy between tunneling systems at
the corresponding temperature. Our results provide clear evidence for the
interaction between tunneling systems, whose energy is in quantitative
agreement with theoretical expectations, and reveal that without the
knowledge of the corresponding strain dependences, the measured
temperature dependences below $\sim 50~$mK of the acoustic properties of
disordered solids are rather meaningless.

\end{abstract}

\maketitle


\section{INTRODUCTION}
The low temperature properties of amorphous and disordered solids are
influenced to a large extent by the interaction of phonons and conduction
electrons with tunneling systems (TS). These TS are low-energy excitations
whose nature still remains a mystery after thirty years of low-temperature
research. At the beginning of the 80's, Francisco de la Cruz contributed
to this research starting successfully in Bariloche with the production of
amorphous superconducting ribbons (La$_{70}$Cu$_{30}$ and
Zr$_{70}$Cu$_{30}$) by melt spinning using an arc furnace \cite{tut}. With
the help of his wife, Maria Elena de la Cruz, Alberto Ridner and one of
the authors of this contribution (P.E.), he directed a research work to
investigate the influence of heat treatments on the thermal conductivity
$\kappa(T)$, among other transport properties, of amorphous
superconductors. Well below the critical temperature $T_c$, the thermal
conductivity is inversely proportional to the density of states $P_0$ of
the TS, multiplied by the coupling constant between TS and phonons
$\gamma$, i.e. $\kappa(T<<T_c) \propto T^2/(P_0 \gamma^2)$. From these
data one obtains the product $P_0\gamma^2 \sim 1~$J/g for amorphous
superconductors in the as-quenched state \cite{esqk}. This value is of the
same order of magnitude as for the amorphous metal PdSiCu and a factor 5
to 10 smaller than for dielectric glasses \cite{esqr}. Low-temperature
annealing of the amorphous metal, relaxing but not crystallizing the
amorphous structure, leads to an increase of $\kappa(T<<T_c)$ as well as
to changes in the acoustic properties \cite{esqk,esqluz}. From these as
well as from specific heat results it was concluded that the observed
changes at the early relaxation states are mostly due to the decrease of
the coupling constant $\gamma$ rather than to a change in the density of
states of TS (see page 194 ff in Ref.~\cite{esqr}).

From measurements done during the last 20 years \cite{esqr}, we learn that
a ``relaxed" amorphous structure may show vanishingly small glasslike
properties, though still being in the amorphous, disordered state. Since
some non amorphous solids show glasslike properties, we conclude that
amorphousness is neither a necessary
 nor a sufficient condition for the existence of glasslike anomalies at
low temperatures. Then, the key question is: where does the
low-temperature ``universality" of glasses rely on? The universality
properly refers to an apparent ``maximum" value observed in {\em bulk}
samples for the ratio $C = P_0\gamma^2 / (\rho v^2) \sim 10^{-3}$ (here
$\rho$ is the mass density and $v$ the sound velocity of the sample),
despite a large variation of all four parameters. The standard tunneling
model, based on the assumption of an arbitrary amount of independent,
noninteracting tunneling entities with a broad distribution of energy and
relaxation time \cite{esqr}, does not provide an answer to this
experimental fact. These assumptions have been questioned \cite{yu}
arguing that the low-energy excitations themselves and the observed
universality are indeed the result of interactions between some kind of
defects.

The acoustic properties in the kHz range are an excellent tool to study
the interaction between phonons and/or electrons with TS. Following the
standard tunneling model\cite{ande72,phil72,phil_rev} (STM), the
measurement of the internal friction $Q^{-1}$ and the relative change of
sound velocity $\Delta v/v$ with temperature provide information on the
density of states of TS ($P_0$) and their coupling to phonons ($\gamma$)
or conduction electrons ($K$). Additionally, the relatively low
frequencies ($\omega$) used enable acoustic measurements to be conducted
at ultralow temperatures ($T < 5~$mK) as the energy dissipation in the
sample ($\propto \omega^3 \,$!) can be kept at extremely small levels ($ <
10^{-13}~$W), thus avoiding self-heating effects. Moreover, this kind of
measurement introduces the applied strain $\epsilon$ as a third variable
parameter. This parameter shows up as important and fundamental as the
temperature to understand the observed deviations of the acoustic
properties from the predictions of the STM.

The result we would like to stress in this contribution is that
the strain dependence of the acoustic properties of some glasses
provides clear evidence for the interaction between TS. The
ubiquitous strain dependence shows up as a clue to understand why
 the temperature dependence of the
acoustic properties of glasses below $\sim 50~$mK is not understood, in
spite of several studies published in the literature during the last
decade on this issue. We will show below that there is a clear difference
between the strain dependence of the acoustic properties obtained for
dielectric and metallic glasses in the superconducting state, in
disagreement with the expectations. In this contribution we will deal with
similar superconducting alloys as those studied by Francisco de la Cruz
(Zr$_{x}$Cu$_{1-x}$) \cite{tut,esqk}, but now with critical temperatures
below 0.3 K (for $x = 0.3, 0.4$).

\section{THEORETICAL BACKGROUND}

\subsection{Predictions of the Standard Tunneling Model}

The STM, independently introduced by Anderson, Halperin and Varma
\cite{ande72}, and Phillips \cite{phil72} in 1972, provides a
phenomenological description of thermal, dielectric and acoustic
properties of amorphous solids below $\sim 1$~K
\cite{esqr,phil_rev,hunk86}. The central idea of the model is the
universal existence in glasses of a random distribution $P
(\Delta_0 , \Delta) = P_0 / \Delta_0 $ of {\em independent}
two-level states or TS with quantum energy splitting $\Delta_0$
and asymmetry $\Delta$. The acoustic properties in the kHz range
are much more sensitive than thermal properties to details of the
TS and their interactions with phonons and/or electrons. At
very-low temperatures and not very-high frequencies ($T \ll T_{\rm
co}$), the STM predicts \cite{esqr,phil_rev,hunk86} for {\em
dielectric} glasses a relative change of sound velocity $\Delta
v/v = C \ln(T/T_0)$ and an internal friction $Q^{-1} \propto C T^3
/ \omega$. Above the crossover temperature $T_{\rm co}$ ($T \gg
T_{\rm co}$), (phonon) relaxational processes add to resonant
contributions to the sound velocity and $\Delta v/v = - (C/2)
\ln(T/T_0)$, whereas the relaxation-dominated internal friction
reaches a constant ``plateau'' value $Q^{-1} = (\pi/2) C$.  $T_0$
is an arbitrary reference temperature. In amorphous {\em metals}
\cite{blac81}, conduction electrons provide in addition to phonons
an alternative channel for relaxation of the TS, typically
dominant below 1 K. In this case, $T_{\rm co} \sim$ 1--10 $\mu $K
is estimated instead of $T_{\rm co} \sim $ 100 mK as for
dielectric glasses, and experiments should therefore show in the
whole measured temperature range a sound-velocity variation with
both resonant and electron relaxational contributions $\Delta v/v
= (C/2) \ln(T/T_0)$, and an internal friction with a wide plateau.

The first measurements of the acoustic properties of an amorphous
superconductor, Pd$_{30}$Zr$_{70}$, in the superconducting {\it and}
normal state, were performed by Neckel et al.\cite{neckel}. Well below its
transition temperature $T \ll T_{\rm c}$, an amorphous superconductor was
expected to behave as a dielectric glass. However, this and subsequent
experiments of the same kind\cite{esqr} have shown that the interaction
mechanisms between TS and conduction electrons are still far from being
well understood.

Although some experiments \cite{esqr,phil_rev,hunk86,rayc84} have given
support to the STM, significant discrepancies below $\sim 100 \,$mK have
been reported more recently: The internal friction in dielectric glasses
decreases for $T < T_{\rm co}$ slower than $Q^{-1} \propto T^3$
\cite{esqr,esqu92,hunk00}, the slopes of $\Delta v/v$ vs. $T$ sometimes
differ from the predicted values, and the acoustic properties of both
dielectric and metallic glasses exhibit a strong, unexpected strain
dependence \cite{esqr,esqu92,ramos00}.

Some of the observed strain dependences can be accounted for by including
the strain energy in the population of the energy levels in equilibrium,
i.e.
\begin{equation}
n_0 \propto \tanh(\sqrt{\Delta_0^2+(\Delta + 2\gamma \epsilon_0
\sin(\omega t))^2}/2k_{\rm B}T) \label{eq1} \end{equation} where
$\epsilon_0$ is the amplitude of the stress field \cite{esqr}. Numerical
results obtained with  values of the strain $\epsilon_0$ similar to those
used in experiments with SiO$_2$ resemble partially some of the measured
features \cite{esqr}, namely: (1) For strain energies of the order of the
thermal energy $\gamma \epsilon_0 \sim k_{\rm B} T$ the sound velocity
decreases with strain and shows a higher slope with $T$ at $T < T_{\rm
co}$, (2) the position of its maximum at $T_{\rm co}$ shifts to higher
temperatures with strain, and (3) at $T \ll T_{\rm co}$ the sound velocity
saturates. The internal friction, on the other hand, is not affected by
the strain at $\epsilon_0 \le 10^{-7}$, whereas it appears to decrease
slightly at higher strains \cite{esqu92}. A more rigorous approach has
been done in Ref.~\cite{sto95} and the results resemble those obtained
using the simple assumption of a population change. In the limit of slow
relaxation of the TS's at $T \ll T_{\rm co}$, analytical results were also
obtained for the change of $Q^{-1}$ with strain \cite{esqr,sto95}.

We note that Stockburger et al.\cite{sto95,sto94} developed a theory based
on the tunneling model, which accounted for a strain dependence of the
sound velocity and the internal friction of amorphous metals \cite{sto94}
and dielectric glasses \cite{sto95}, in terms of an intrinsically
nonlinear response of the TS to an external disturbance. This saturation
regime is achieved when $\gamma \epsilon_0 > k_{\rm B} T$. In this work,
however, we will deal always with applied strains orders of magnitude
below this saturation limit.

\subsection{Ensemble of Interacting Tunneling Systems}

As pointed out above, one striking feature of disordered solids is the
universality of the maximum value of the plateau observed in the internal
friction of bulk dielectric glasses $Q^{-1}_{\rm plateau} \sim 10^{-3}$,
despite a great variation of the parameters involved. As a possible reason
for this fact, Yu and Leggett \cite{yu} argued that this may be the result
of interactions between some kind of defects. On general grounds
\cite{yu}, the effective interaction between two TS separated by a
distance $r$ is dipolar elastic and the interaction energy reads $\sim U_0
/ r^3$, with $U_0 \approx \gamma^2 / \rho v^2$. The observed universality
of low-temperature glassy properties can therefore be written in terms of
the dimensionless interaction strength as $ P_0 U_0 \approx {\rm const} \,
\approx C \sim 10^{-3}$ \cite{burin}.

In a recent paper, Burin et al. \cite{burin01} considered the
delocalization of low-energy excitations in disordered systems due
to an external alternating acoustic or electric field. The
delocalization depends on both the strain amplitude and the
distance between resonant TS neighbors, which depends on the
elastic interaction $U_0$ between them. Similarly to
Eq.~(\ref{eq1}) the asymmetry $\Delta$ becomes strain and time
dependent and may become smaller than $\Delta_0$ for large enough
strain amplitudes. In this case, they argue that for sufficiently
small frequencies compatible with our acoustic frequencies, the
strain {\em increases} the effective number of delocalized TS, {\em
increasing} $Q^{-1}$ below $T_{\rm co}$ due to a shift in  the
crossover condition $\omega \tau < 1$ to lower temperatures,
$\tau$ being the characteristic relaxation time of the TS.
Recently done measurements of the dielectric constant of an
insulating glass as a function of electric field and thickness
apparently show agreement with these theoretical predictions
\cite{ladieu}. However, the predicted changes due to
delocalization with strain appear to be at odds with the changes
with strain observed in the sound velocity and internal friction
of SiO$_2$ \cite{esqu92}. We will see in the next section that the
changes with strain observed in amorphous metals also disagree
with these predictions. We will consider and discuss the role of
the strain energy within the model of interacting TS pairs using a
different point of view in the next section, after presenting our
experimental results.

\section{EXPERIMENTAL RESULTS AND DISCUSSION}

Ribbons of amorphous Zr$_{\rm x}$Cu$_{\rm 1-x} (\sim 30 \mu $m thick) were
prepared with the melt-spinning technique in an argon-controlled
environment similar to that described in Ref.~\cite{tut}. Young-modulus
sound velocity variation and acoustic attenuation were measured at around
1 kHz using the vibrating-reed technique \cite{esqr,rayc84} in a nuclear
adiabatic demagnetization cryostat \cite{gloos}. The measurements were
made as a function of temperature (0.1~mK$~\le T \le~1$~K), both in the
superconducting and normal state of the samples ($T_c \simeq 280$~mK and
95~mK for $x = 0.4$ and 0.3, respectively), and also as a function of the
applied strain ($4 \times 10^{-11} \lesssim \epsilon \lesssim 10^{-8}$) at
constant temperatures. For the measurements as a function of temperature,
the driving voltage remained constant (and not the applied strain).
\begin{figure}
\vspace{-0.3cm}\centerline{\psfig{file=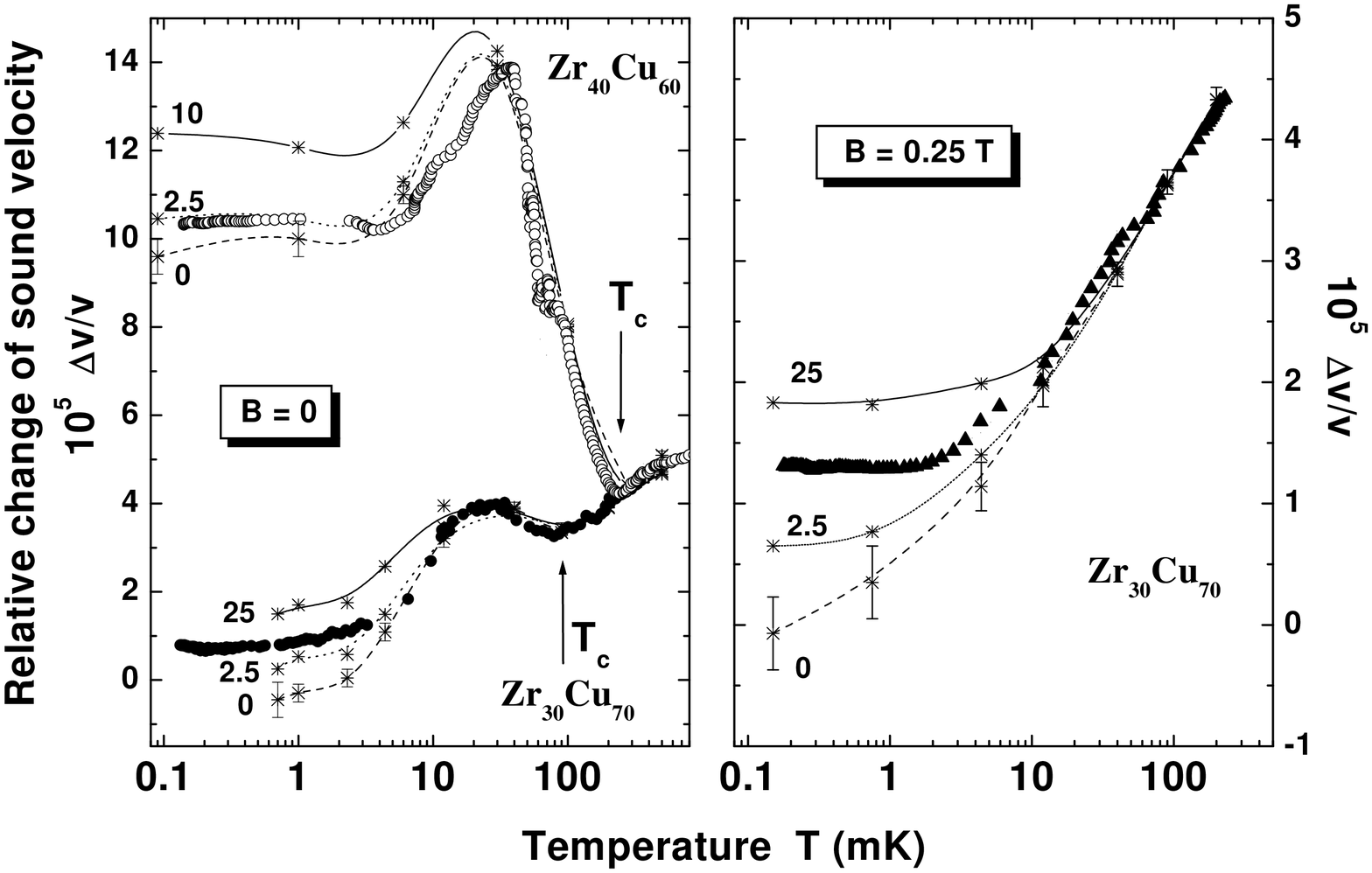,height=3.5in}}
\caption{Left panel: Sound-velocity variation of amorphous superconductors
Zr$_{30}$Cu$_{70}$ (solid symbols) and Zr$_{40}$Cu$_{60}$ (open symbols),
measured at 1 kHz with a driving voltage of 1 V. The corresponding
superconducting transition temperatures $T_c$ are indicated by arrows.
Right panel: Sound-velocity variation of Zr$_{30}$Cu$_{70}$ in normal
metallic state at a driving voltage of 2 V. In all cases, extrapolated
data to zero strain and interpolated data for constant-strain curves (see
text) are also shown, with lines as eyeguides: labels indicate the
corresponding strains in units of $10^{-10}$.} \label{fig:1} \end{figure}

In order to study the effects of the applied strain in greater detail, we
conducted\cite{koenig02} complementary experiments by carefully measuring,
at selected constant temperatures, the full resonance curves for several
given excitation voltages, obtaining the corresponding applied strain from
the reed vibration amplitude\cite{ramos00}. It is important to stress that
each plot of $\Delta v/v$ and $Q^{-1}$ as a function of strain consists of
up to 30 data points, each of which is the result of a single resonance
curve measured in average in one hour. This implies that the temperature
had to be kept constant for at least 24h even at the lowest temperatures,
far below 1mK. Following this procedure, we were able to obtain curves of
as a function of the applied strain (see Figs. 4 and 5 in
Ref.~\cite{ramos00} for PdSiCu and Fig.~2 in Ref.~\cite{koenig02} for
Zr$_{30}$Cu$_{70}$) for several selected temperatures. By interpolating
those curves we can obtain true {\it constant-strain} curves of the
acoustic properties as a function of temperature, and even their presumed
very-low-strain behavior by {\it extrapolating} them to zero.

\begin{figure}
\centerline{\psfig{file=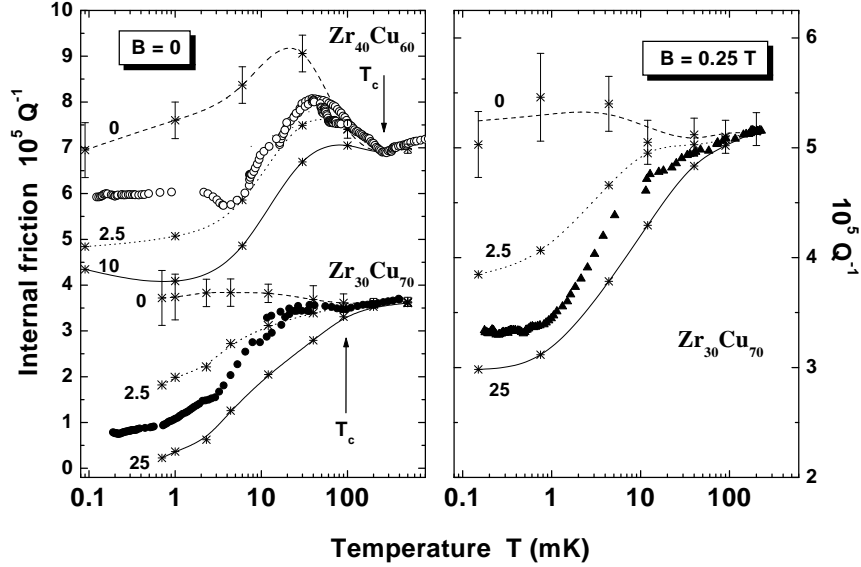,height=3.5in}} \caption{Left
panel: Internal friction of amorphous superconductors Zr$_{30}$Cu$_{70}$
(solid symbols) and Zr$_{40}$Cu$_{60}$ (open symbols), measured at 1 kHz
with a driving voltage of 1 V. The corresponding superconducting
transition temperatures $T_c$ are indicated by arrows. For clarity the
data for Zr$_{30}$Cu$_{70}$ have been shifted vertically by $-1.5 \times
10^{-5}$. Right panel: Internal friction of Zr$_{30}$Cu$_{70}$ in normal
metallic state at a driving voltage of 2 V. In all cases, extrapolated
data to zero strain and interpolated data for constant-strain curves (see
text) are also shown, with lines as eyeguides: labels indicate the
corresponding strains in units of $10^{-10}$.} \label{fig:2}
\end{figure}

 In Fig.~\ref{fig:1} and Fig.~\ref{fig:2}, we show
$\Delta v/v$ and $Q^{-1}$ data, respectively,  for Zr$_{30}$Cu$_{70}$, in
both superconducting (SC) and normal (N, with a magnetic field $B =
0.25$~T, which is above the critical field) states, obtained at a given
excitation voltage of the reed during warm-up periods of the nuclear
stage. Similar data were obtained for Zr$_{40}$Cu$_{60}$ (with $B =
0$)\cite{koenig02,koenigB}, that are also shown in the left panels of
Figs.~\ref{fig:1} and~\ref{fig:2}, as well as for the metallic glass
PdSiCu\cite{ramos00}. Within the STM, the maximum in $\Delta v/v$ at
$T_{\rm co} \approx 30 $~mK and the corresponding decrease of $Q^{-1}$
with decreasing temperature should originate from the abovementioned
crossover, occurring when $\omega \tau \sim 1$. As expected, this peak
disappears in the normal metallic state, since this crossover shifts to
orders of magnitude lower temperatures due to the much faster relaxation
rate of TS by electrons than by phonons.

However, at lower temperatures both $\Delta v/v$ and $Q^{-1}$ strongly
depend on the excitation voltage (hence on strain), a behavior similar to
that observed in the metallic glass PdSiCu\cite{esqu92,ramos00}. The sound
velocity clearly increases and the internal friction decreases with
increasing voltage or strain, and furthermore they do not exhibit in any
case below $\approx 10 $~mK the behavior predicted by the STM. Moreover,
we remind that the STM is linear in strain and cannot account for any
strain dependence of acoustic properties. Some of the constant-strain
curves are also shown in Fig.~\ref{fig:1} and Fig.~\ref{fig:2}. We note
that the interpolated constant strain curves for $\epsilon > 10^{-10}$
confirm the main temperature dependence observed from direct
constant-voltage curves.

 The following facts are interesting to note: (a)
The apparent crossover at $T_{\rm co} \sim 30~$mK observed in
Zr$_{30}$Cu$_{70}$ in the superconducting state is an artifact due to the
applied strain. The internal friction does not start to decrease at this
temperature but it remains in first approximation independent of $T$ for
low enough strains. (b) The conduction electrons influence both acoustic
properties and produce the anomaly at $T_c$, clearly recognized after the
application of the magnetic field and in good agreement with previous
measurements at higher temperatures\cite{esqluz,neckel}. However, two new
results are obtained from the strain dependence: (c) At $T < 2~$mK the
sound velocity shows similar $T-$dependence both in the normal and
superconducting state. (d) The strain dependence of the acoustic
properties (in particular $Q^{-1}$) depends on the state of the sample.
This, as well as the results to be presented in Fig.~3, indicate that the
conduction electrons influence the strain dependence \cite{koenig02}. All
these results cannot be explained within the STM. \vspace{-2.5cm}

\begin{figure}
\vspace{-1.cm}\centerline{\psfig{file=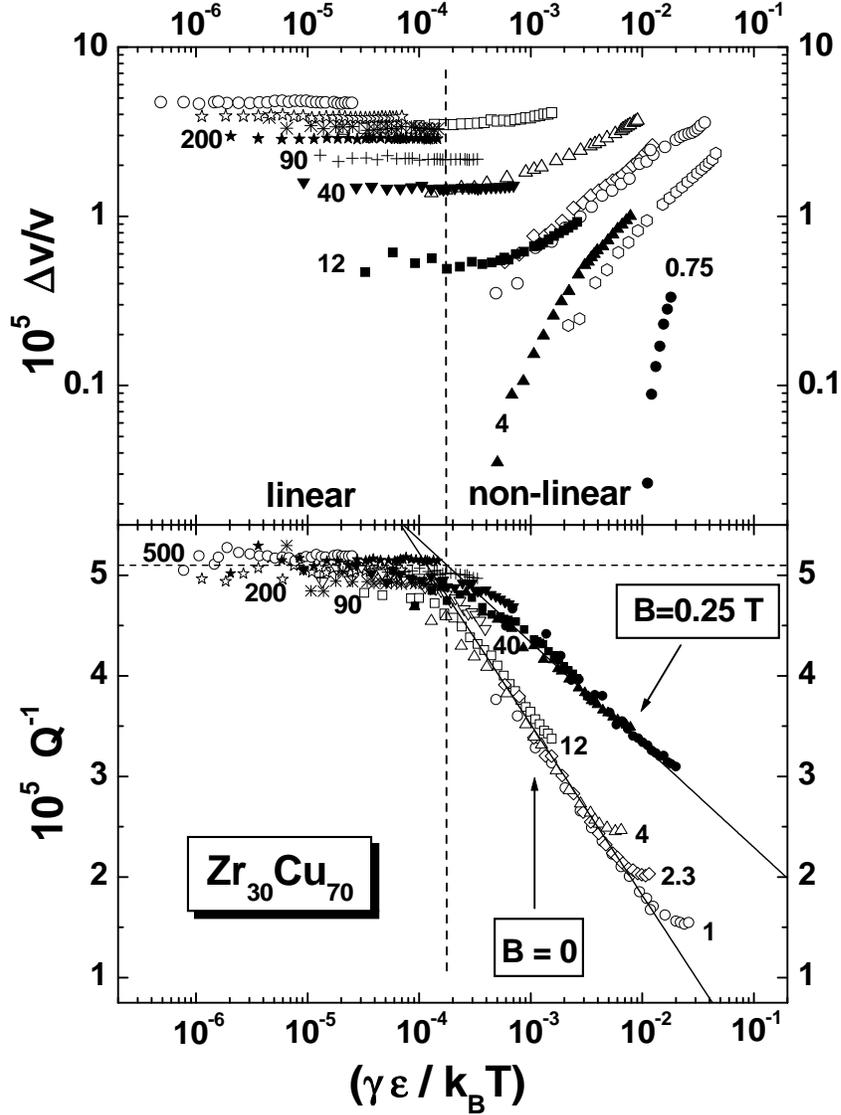,width=4.5in}}
%
\caption{Sound-velocity variation $\Delta v/v$ (top) and internal
friction $Q^{-1}$ (bottom) measured for Zr$_{30}$Cu$_{70}$  (open
symbols, $B = 0$; solid symbols, $B = 0.25$ T), plotted versus the
ratio of strain energy to thermal energy in a logarithmic scale,
for several sets of constant-temperature data which are labelled
in mK. Solid lines show the linear decrease of $Q^{-1}$ with
log($\gamma \epsilon / (k_{\rm B} T)$) above the crossover at
$\gamma \epsilon / (k_{\rm B} T) \approx P_0 U_0$.} \label{fig:3}
\end{figure}

\vspace{+2.0cm} In Fig.~\ref{fig:3}, all sets of constant-temperature
measurements for Zr$_{30}$Cu$_{70}$ are plotted as a function of the
dimensionless ratio of strain energy to thermal energy, $\gamma \epsilon /
k_{\rm B} T$, along almost five orders of magnitude. A typical value of
$\gamma = 1 $~eV has been used for the coupling constant. As can be
observed, all curves show the same $Q^{-1}$ ``plateau'' value at $\gamma
\epsilon / k_{\rm B} T \, < \, 10^{-4}$ (strikingly, also in the SC,
dielectric case). Moderately increasing the strain, so that $\gamma
\epsilon / k_{\rm B} T \, > \, 10^{-4}$, makes $Q^{-1}$ to decrease
logarithmically with $\gamma \epsilon / k_{\rm B} T $ (the slope of the
solid line shown in Fig.~\ref{fig:3} is $\log( \gamma \epsilon / k_{\rm B}
T ) / 3 $) for the SC case. When conduction electrons contribute to the
relaxation of TS in the whole experimental temperature range, the same
behavior is observed, though the slope is less steep ($ \log( \gamma
\epsilon / k_{\rm B} T ) / 5 $). The difficulties encountered in
zero-strain extrapolations at the lowest temperatures (say, below 5--10
mK) is also made clearer now: even the very low strains used here are not
low enough to completely reach the ``plateau'' limit in some cases. On the
other hand, $\Delta v/v$, which is rather governed by resonant processes
(especially in the SC state), also exhibit a crossover at $\gamma \epsilon
/ k_{\rm B} T \, \sim \, 10^{-4}$ from the strain-independent or linear
behavior into a regime of strong increase of the sound velocity with
strain, either in N or SC states. Furthermore, although we have shown and
discussed here only the case of Zr$_{30}$Cu$_{70}$, a very similar
plot\cite{koenig02} can be obtained for Zr$_{40}$Cu$_{60}$ data in the SC
state, and also for the normal metal PdSiCu, pointing once again to a
universal behavior of glasses.

We want to emphasize that in all cases the crossover from
strain-independent to strain-dependent behavior occurs when the above
defined ratio is $\gamma \epsilon / k_{\rm B} T  \approx $ 2--5
$Q^{-1}_{\rm plateau} \approx  P_0  U_0$. Therefore, one can make the
following argumentation within the interacting model: $P_0 \times k_{\rm
B} T$ is the number of thermally activated TS per unit volume, then $P_0
U_0 \times k_{\rm B} T$ is the total effective interaction energy between
TS at the considered temperature. We thus have two limits taking
into account the strain energy:\\
(i) For $\gamma \epsilon \ll ( P_0 U_0 \times k_{\rm B} T )$, i.e.
when the strain energy driven by the external sound wave is less
than the interaction energy between TS, one may observe the
``true'' unperturbed response of the TS, in which only their
interaction and the thermal energy determine the temperature
dependence of the acoustic properties. Since in most of the
experiments performed in the past the strain energy was not
considered, we do not know a priori if the STM applies at this
limit. Going a step further, we may speculate that in this regime
the TS are delocalized in the sense that their interaction energy
overwhelms all other perturbations (thermal and of the applied
strain) and coherent behavior is expected. In this case we may
have the maximum
possible density of TS as scattering centers for phonons.\\
(ii) When $\gamma \epsilon \gg ( P_0 U_0 \times k_{\rm B} T )$, the system is
perturbed and nonlinear effects appear in the acoustic properties. In this
case the linear STM picture of independent TS is surely no longer valid. Moreover, we
expect that the strain energy influences the coupling between TS in such a
way that an effective localization and reduction of the density of states
may arise, since a ``localized" TS has less probability to scatter
resonantly a phonon than in the case of an ensemble of TS with a
broader spectrum of energies and relaxation times.

In other words, the ratio $\gamma \epsilon / (k_{\rm B} T)
\thickapprox P_0 U_0 \sim Q^{-1}_{\rm plateau}$ would therefore
indicate the border between a linear and a nonlinear regime given
by the competition between interaction and perturbation energies.
This crossover will occur around $\sim $ 20--50~mK for typical
experiments. Strictly speaking, it depends on the variable $\gamma
\epsilon / k_{\rm B}T$, since strain and thermal energies play an
inextricably combined role in the acoustic properties of glasses
at low temperatures due to the mutual interaction of TS.

Interestingly, the reported behavior on the acoustic properties of these
metallic glasses, either in N or SC state, as a function of the strain, is
very different from that found \cite{esqu92} in vitreous silica,
v-SiO$_2$, where the sound velocity {\it decreases} and the maximum in
$\Delta v/v$ shifts to higher temperatures with increasing strain
$\epsilon$ varying from $2 \times 10^{-8}$ to $10^{-6}$, whereas the
internal friction $Q^{-1}$ exhibits a negligible strain dependence. Since
in our work on metallic glasses we have employed a very low strain range
(at least two orders of magnitude lower than those used in the SiO$_2$
experiments), it might be that the aforementioned effect of the strain
energy changing the thermal population of the energy levels in equilibrium
(see Eq.~(\ref{eq1})) is dominant in that higher strain range used for
v-SiO$_2$. In addition, at the highest applied strains for the lowest
temperatures measured in Zr$_{\rm x}$Cu$_{\rm 1-x}$, i.e. for $\gamma
\epsilon / k_{\rm B} T > 10^{-2}$ (see Fig.~\ref{fig:3}), the logarithmic
decrease of the internal friction with increasing $\gamma \epsilon /
k_{\rm B} T$ seems ceasing to hold. So, both effects altogether could
explain the observed strain-dependence behavior of v-SiO$_2$. Another
speculative possibility is that vitreous silica exhibits indeed a behavior
as a function of strain opposite to other glasses, as it does in other
physical properties due to its particular tetrahedral network.

In summary, we have presented a detailed study of the acoustic  properties
of some metallic glasses in the temperature range 0.1~mK$~\le T \le~1$~K,
in both N and SC states. The main experimental finding has been a general,
strong strain dependence of the acoustic properties of glasses observed at
very low temperatures in all the studied cases, for both internal friction
and sound velocity variation. In the normal metallic state, sound-velocity
variation $\Delta v/v$ and internal friction $Q^{-1}$ seem to recover the
expected STM behavior in the zero-strain limit. Nevertheless, in the
superconducting state the plateau in $Q^{-1}$ also extends down to the
lowest temperatures ($T < 1$~mK) in the zero-strain limit instead of
exhibiting the expected $\propto T^3$ dependence, predicted by the STM.

By deeper analyzing all these data, we have found a  general crossover
from a linear to a non-linear regime with increasing the ratio of strain
over thermal energy. In fact, this crossover occurs always when $\gamma
\epsilon / k_{\rm B} T \, \approx \, P_0  U_0$, in agreement with the
interaction energy postulated by Yu and Leggett \cite{yu}, and by Burin
and Kagan\cite{burin}. Furthermore, our results show that the measured
$T$-dependences of the acoustic properties of disordered materials below
$\sim $~50~mK are rather meaningless, unless one measures the
corresponding strain dependences too. We have also shown that none of the
current models in the literature are able to account for the temperature--
and strain--dependent behavior of the acoustic properties of amorphous
metals in the {\it superconducting} state at very low temperatures.

\section*{ACKNOWLEDGMENTS}
This research was supported by the DFG, DAAD and EU Contract ERB FMGE
CT950072. Fruitful discussions with many participants of the Workshop  on
Collective Phenomena in the Low-Temperature Physics of Glasses (Dresden,
2003) are gratefully acknowledged.

\end{document}